\documentclass[aps,prl,twocolumn,superscriptaddress,showpacs]{revtex4}
\usepackage{graphics}
\usepackage{subfigure}
\usepackage{times}

\def\tlfs{TlFe$_{2-x}$Se$_2$}
\def\kfs122{KFe$_2$Se$_2$}
\def\kfsx22{K$_{1-x}$Fe$_2$Se$_2$}

\def\bfa122{BaFe$_2$As$_2$}
\def\fs11{FeSe}

\begin{document}

\preprint{}
\title{Electronic Structure and Mott Localization in Iron
Deficient TlFe$_{1.5}$Se$_2$ with Superstructures}

\author{Chao Cao}
  \affiliation{Condensed Matter Physics Group,
  Department of Physics, Hangzhou Normal University, Hangzhou 310036, China}

\author{Jianhui Dai}
  \affiliation{Condensed Matter Physics Group,
  Department of Physics, Hangzhou Normal University, Hangzhou 310036, China}
  \affiliation{Department of Physics, Zhejiang University, Hangzhou 310027, China}

\date{\today}

\begin{abstract}
Electronic structure and magnetic properties for iron deficient
TlFe$_{2-x}$Se$_2$ compounds are studied by first-principles
calculations. We find that for the case of $x=0.5$ with a Fe-vacancy
ordered orthorhombic superstructure, the ground state exhibits a
stripe-like antiferromagnetic ordering and opens a sizable band gap
if the short-ranged Coulomb interaction of Fe-3d electrons is
moderately strong, manifesting a possible Mott insulating state.
While increasing Fe-vacancies from the $x=0$ side, where the band
structure is similar to that of a heavily electron-doped \fs11
system, the Mott localization can be driven by kinetic energy
reduction as evidenced by the band narrowing effect. Implications of
this scenario in the recent experiments on TlFe$_{2-x}$Se$_2$ are
discussed.
\end{abstract}


\maketitle 

The discovery of superconductivity (SC) with critical temperatures
up to 56 K in iron pnictides
\cite{lofa_discovery,xhchen_nature_453_761,nlwang_prl_100_247002,zxzhao_epl_83_17002,hhwen_epl_82_17009,cwang_epl_83_67006}
has triggered renewed interest in searching new route to
high-temperature SC. Considerable concerns have been focused on the
nature of the parent compounds which show variable bad metal
behavior\cite{qazilbash_np_5_647} and a universal strip-like
antiferromagnetic (SDW) order\cite{pcdai_nature_453_899}. The
magnetic ordering was proposed to be the consequence of the low
energy states ( or itinerant electrons ) within the nearly nested
Fermi surfaces, and hence the Fermi surface nesting is responsible
to SC when the SDW order is suppressed
\cite{dong_epl_83_270069,singh_1111,mazin_1111}. An alternative
possibility is that the magnetic structure is due to the strong
correlation among the Fe-3d electrons, so that the states far away
from the Fermi surfaces should be taken into account as well. To
this end, Fermi surface nesting is not a necessary ingredient and
the $J_1$-$J_2$ Heisenberg model based on the local moment picture
could be an appropriate starting point.
\cite{yildirim_1111,qsi_prl_101_076401,cao_1111,zylu_prb_78_2245179}

Most recently, Fang {\it et al.} \cite{mhfang_1012} reported that
the Fe-deficient compounds (Tl,K)Fe$_{2-x}$Se$_2$ exhibit SC with
$T_c$ up to $\sim$ 31 K for $x=0.12\sim 0.3$. Special interest in
this class of materials is that the SC emerges in proximity to an
insulating phase. It is yet unknown whether this insulating phase is
a Mott insulator driven by Fe-3d electron correlations.
Theoretically, this possibility is mysterious, as previous
first-principle calculations on both TlFe$_2$Se$_2$
\cite{singh_prb_79_094528} and KFe$_2$Se$_2$
\cite{Shein,cao_dai}with the ThCr$_2$Si$_2$ structure (122-type)
suggest that the parent compounds of the ternary iron chalcogenides
should be metallic with either checkerboard antiferromagnetic (AFM)
(for Tl-122)\cite{singh_prb_79_094528} or SDW (for
K-122)\cite{cao_dai} order, much like the electron overdoped 11-type
iron selenides. Experimentally, it has been reported recently that
the alkali intercalated compounds
K$_{0.8}$Fe$_2$Se$_2$\cite{xlchen_prb_82_180520} and
Cs$_{0.8}$Fe$_2$Se$_{1.96}$ \cite{luetkens} ( both iso-structural to
\bfa122), while superconducting under 30 K and 27 K respectively,
exhibit the metallic behavior in their normal states.

Two closely related questions thus arise: What is the ground state
of the "parent" compound of these superconducting ternary iron
chalcogenides of 122-type structure, and how can a Mott-insulating
phase develop and then diminish with increasing Fe-content  or {\it
electron doping} giving way to SC?

In this paper, we suggest partial answers to the questions by
first-principles study on TlFe$_{2-x}$Se$_2$. We start from and pay
special attention to the case with $x=0.5$, which is
stoichiometrically equivalent to Tl$_2$Fe$_{3}$Se$_4$ but with the
122-type structure. Hence, Fe$^{2+}$ is the nominal valence as in
other iron pnictides/chalcogenides. Our calculation shows that the
ground state of TlFe$_{1.5}$Se$_2$ with Fe-vacancy ordered
orthorhombic superstructure is an SDW phase which can open a sizable
gap if a moderately strong electron correlation is imposed. The
quantitative change of the bandwidth of Fe-3d electrons further
supports the Mott localization driven by kinetic energy reduction
due to Fe-vacancies. This scenario is quite similar to the Mott
localization proposed for iron oxychalcogenides
La$_2$O$_2$Fe$_2$O(Se,S)$_2$\cite{jxzhu_prl_104_216405}, where the
reduction of kinetic energy or the band narrowing is due to the
expanded interatomic Fe-Fe distance. We will discuss the
implications of this scenario in the recent experiments on
TlFe$_{2-x}$Se$_2$.

The electronic structure calculations were performed with the Vienna
Ab-initio Simulation Package (VASP)\cite{vasp_1,vasp_2}. All
structures were optimized so that the forces on individual atoms
were smaller than 0.02 eV/\AA\ and the pressure convergence
criterion is chosen to be 0.5 kbar. For the optimization and ground
state calculations, a $8\times 4\times 4$ Monkhorst-Pack
k-grid\cite{mp_kgrid} was employed, while $16\times 8\times 8$
Monkhorst-Pack k-grid was used for the density of states (DOS)
calculations. The PBE flavor of general gradient approximation (GGA)
to the exchange-correlation functional\cite{PBE_xc} was applied
throughout the calculations.

\begin{figure}[htp]
 \centering
 \scalebox{0.15}{\includegraphics{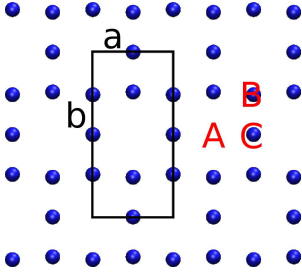}}
 \caption{Superstructure of TlFe$_{1.5}$Se$_2$. $A$, $B$, and $C$ correspond
 to three different sites directly above the Fe-vacancy, the 3-coordinated
Fe site, and the 2-coordinated site, respectively. Another stacking
pattern $A'$ is that the upper-layer is rotated 90$^\circ$ around
$c$-axis.
   \label{fig_structure}}
\end{figure}

We first discuss the possible crystal structure and spin
configurations of TlFe$_{1.5}$Se$_2$. The M\"{o}ssbauer experiment
suggests a body-centered orthorhombic (BCO) crystal structure
\cite{Seidel}, where the Fe sheets form superstructure due to
Fe-deficiency (FIG. \ref{fig_structure}). There are four possible
stacking configurations for the two neighbouring
Fe-layers\footnote{\label{stacking_arg} We find that the
Fe-deficiency will induce structural relaxation in general, which
leads to the distortion of the square Fe lattice and the small
displacement of Fe atoms. Thus the $AA'$-stacking is energetically
unfavorable and is not considered here.}, as shown in FIG.
\ref{fig_structure}. We consider the AFM and SDW configurations
within each layers, while the interlayer coupling along c-axis can
be either ferromagnetic (FM) or AFM. The relative energies for
different stacking and ordering patterns are listed in Table
\ref{tab_energy}. We find that the ground state ( for $x=0.5$ )
should be of essentially two-dimensional SDW, as the energy
differences among the different inter-layer magnetic orderings are
very small ($<5$meV/Fe), indicating the negligible inter-layer
magnetic coupling at this state. Hence, without losing generality,
we shall focus on the $AA$-stacking pattern in the following
discussions.

We also notice that for the NM state, the full structural
optimization leads to the well-known $c$-collapse problem; while for
the SDW state, the resulting structure ($c$ and $z_{\mathrm{Se}}$)
is within the expectation. Thus, unless otherwise specified, the
electronic structure calculations in the following discussions were
performed with the geometry relaxed in the SDW configuration. The
calculated local magnetic moment of Fe atom is then
$m_{\mathrm{Fe}}$ $\approx$ 2.6$\sim$2.7 $\mu_B$.

\begin{table}[htp]
 \caption{Configuration energies of TlFe$_{1.5}$Se$_2$ ( in unit of meV ).
 $AA$, $AB$, $AC$ and $AA'$ indicates the stacking geometry,
while NM, AFM, SDW indicate the intralayer magnetic configurations.
The superscripts $a$ and $f$ indicate interlayer AFM and FM
orderings.}
 \label{tab_energy}
 \begin{tabular}{c|c|c|c|c}
     &  $AA$ &  $AB$ &  $AC$ & $AA'$ \\
 \hline
  NM  &  0.0  &  9 &  3 &  12.7  \\
 AFM$^f$ & -200   & -177  &  -172  & \\
 AFM$^a$ & -173   & -177   & -174   & \\
 SDW$^f$ & -268   & -268   & -270 & \\
 SDW$^a$ & {\bf -273}   & -268   & {\bf -269}  & \\
 \end{tabular}
\end{table}

\begin{table}[htp]
 \caption{Geometry and magnetic properties of TlFe$_{1.5}$Se$_2$.
 Only the ground state (SDW$^a$) configurations are listed.
$z_{\mathrm{Se}}$ is the internal coordinate for Se and
$m_{\mathrm{Fe}}$ is the local magnetic moment on Fe atoms. The
numbers outside and inside the brackets for $m_{\mathrm{Fe}}$ are
for 3-coordinated and 2-coordinated Fe atoms, respectively.} 
 \label{tab_structure}
 \begin{tabular}{c|c|c|c|c}
     &  &  $AA$ &  $AB$ &  $AC$  \\
 \hline
 SDW$^a$ & a                 & 5.4824 & 5.5407 & 5.5483 \\
         & b                 &11.0519 &11.0497 &11.0270 \\
         & c                 &13.8162 &13.7892 &13.8176 \\
         & $z_{\mathrm{Se}}$ & 0.351  & 0.351(0.350)& 0.350 \\
         & $m_{\mathrm{Fe}}$($\mu_B$) & 2.60(2.73) & 2.61(2.74) & 2.62(2.74	) \\
 \end{tabular}
\end{table}

\begin{figure}[htp]
 \centering
 \subfigure[$U$-dependent DOS]{
  \rotatebox{270}{\scalebox{0.48}{\includegraphics{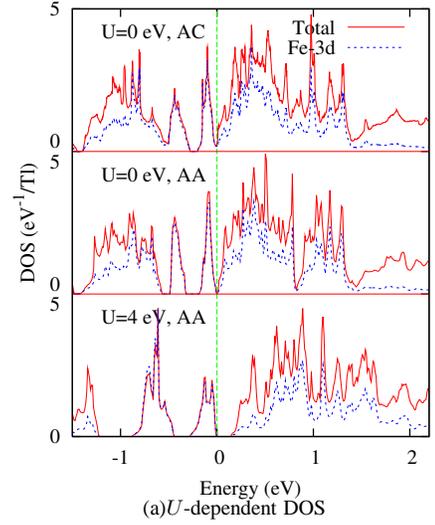}}}
  \label{fig_dos_u}
 }
 \subfigure[$x$-dependent DOS]{
  \rotatebox{270}{\scalebox{0.48}{\includegraphics{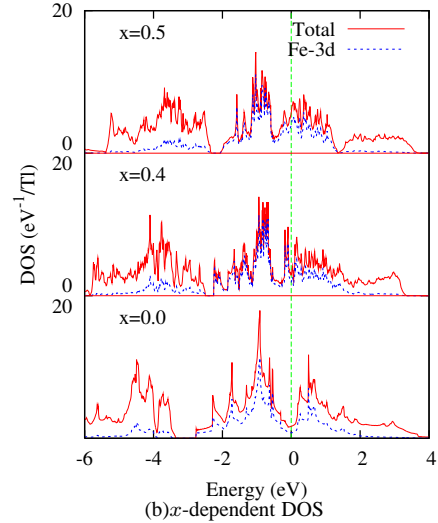}}}
  \label{fig_dos_nm}
 }
 \caption{Total and projected density of states of
\tlfs. \ref{fig_dos_u}: The $U$-dependent DOS at the SDW state for
$x=0.5$; \ref{fig_dos_nm}: The $x$-dependent DOS at the NM state for
$U=0$ with Fe-vacancy superstructures. In \ref{fig_dos_u}, only $\alpha$-spin DOS is shown
since both spin are degenerate. All plots are renormalized to per
~\tlfs~ formula.
   \label{fig_dos}}
\end{figure}

We now examine the DOS of the $AA$-stacking SDW$^a$ plotted in Fig.
\ref{fig_dos_u}, which exhibits a sharp dip around the Fermi energy.
We find that the band gap $E_g$ is vanishingly small for the fully
optimized system under the GGA method. Furthermore, the $AB$- and
$AC$-stacking SDW$^a$ states, which are almost degenerate to the
$AA$-stacking SDW$^a$, are both metallic, with 0.216 states/(eV$\cdot$Formula) and 0.656 states/(eV$\cdot$Formula) respectively. Hence, the three stacking
patterns may coexist at low temperatures. The SDW band gap, if
exists, is too small to compare with the activation gap $E_a\sim
57.7$ meV observed in the $x=0.5$ sample \cite{mhfang_1012}. It
indicates that the observed activation gap in \tlfs ~( at least for
$x=0.5$ ) is not due to the SDW ordering itself.

Here, we suggest that the sizable activation gap can be attributed
to the Mott localization driven by the moderately strong electron
correlation. To seek for this possibility, we extended our
calculations by using the GGA+$U$ method.
The calculated DOS for $U=0$ eV and $U=4.0$ eV are compared in FIG.
\ref{fig_dos_u}. Within GGA+$U$, an insulating gap develops
immediately with increasing $U$, which turns out to be 40, 60, 80,
140, and 230 meV for $U$=1,2,3,4, and 5 eV, respectively. The
projected DOS associated with Fe-3d electrons move to higher
energies mainly distributed around 1 eV above the Fermi energy ( for
 $U$=4 eV ).

\begin{figure}[htp]
  \scalebox{0.248}{\includegraphics{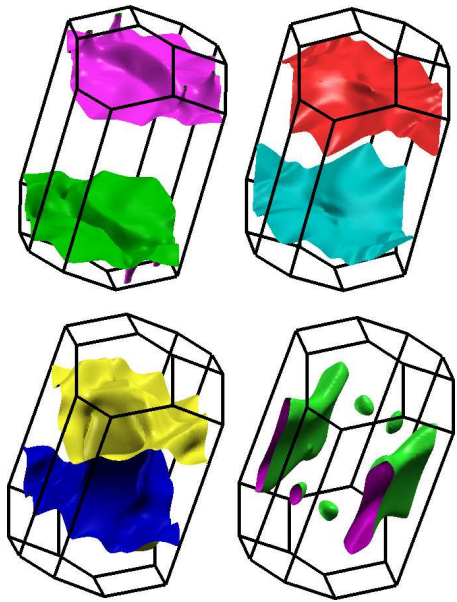}}
 \caption{The Fermi surfaces ( for $x=0.5$ and $U=0$ ) reconstructed
 using the MLWFs shown in the $\Gamma$-centered reciprocal BCO
lattice.
   \label{fig_fs}}
\end{figure}

\begin{figure}[htp]
 \centering
 \subfigure[$x=0.0$]{
  \rotatebox{270}{\scalebox{0.6}{\includegraphics{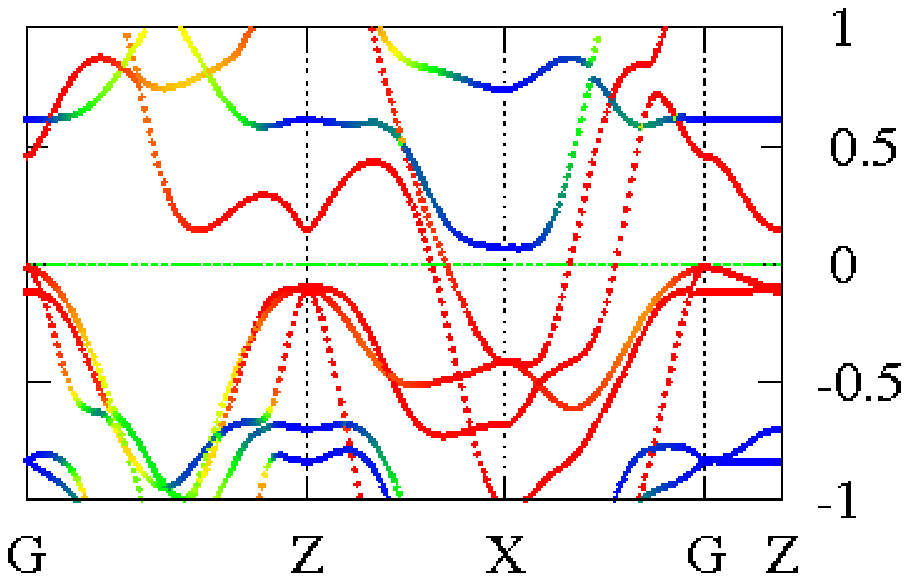}}}
  \label{fig_bands_122}
 }
 \subfigure[$x=0.5$]{
  \rotatebox{270}{\scalebox{0.6}{\includegraphics{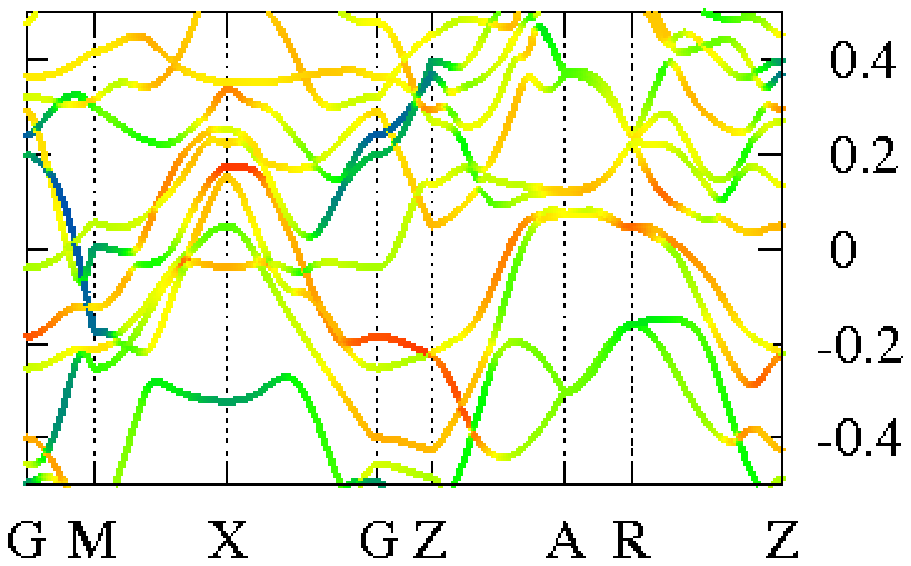}}}
  \label{fig_bands_234}
 }
 \caption{Band structures of \tlfs ~fitted using MLWF.
   \label{fig_bands}}
\end{figure}

We also performed calculations for $x=0,0.4$, keeping the Fe-vacancy
ordered tetragonal superstructure in the latter case
\cite{Seidel,mhfang_1012}. For $x=0$ the AFM state becomes more
stable than the SDW state ( in agreement with
Ref.\cite{singh_prb_79_094528} ), while the overall band structure
 and Fermi surface ( FS ) of TlFe$_2$As$_2$ are similar to that of KFe$_2$Se$_2$
\cite{cao_dai}. For illustration, the band structures for $x=0, 0.5$
in the NM states are fitted to a tight-binding model Hamiltonian by
using the maximally localized wannier function
(MLWF)\cite{mlwf_1,mlwf_2} method. The results are plotted in Fig.
\ref{fig_bands}, where the color indicates the percentage of
d$_{zx(y)}$ and d$_{y^2-x^2}$ composition, from 0\% (blue) to 100\%
(red). For $x=0$, the Hamiltonian is particularly simple with
5-bands. The nearest neighbor hopping parameters and the on-site
energies are listed in TAB. \ref{tab_ham}.  The bands near $E_F$ are
dominated by d$_{zx(y)}$ and d$_{y^2-x^2}$ orbitals except for one
empty band around X. For $x=0.5$, the Hamiltonian is too complicated
due to the structural distortion, and all 5 d-orbitals are entangled
considerably around $E_F$. Nevertheless, the FS of
TlFe$_{1.5}$Se$_2$ can be reconstructed as shown in Fig.
\ref{fig_fs}.

The apparent different band structures for $x=0$ and $0.5$, together
with the drastic change in their FS topologies, provide an
indication for the transition at certain $0<x_c<0.5$, i.e., when
$x>x_c$, the Mott localization takes place. As the Fe-Fe distance
increases less than 1\% from $x=0$ to $x=0.5$ \cite{mhfang_1012},
this amount of lattice expansion is not sufficient for the Mott
localization in \tlfs, as compared to iron oxychalcogenides
La$_2$O$_2$Fe$_2$O(Se,S)$_2$\cite{jxzhu_prl_104_216405}. Here, we
argue that the kinetic energy reduction caused by the Fe-vacancies
should play an crucial role in driving the system to the insulating
phase. An intuitive estimate is that the coordinate number of Fe in
the TlFe$_{1.5}$Se$_2$ is reduced to 3 or 2 depending on the Fe site
comparing to 4 in a perfect square lattice. Thus the total kinetic
energy is substantially reduced by the Fe-vacancies, enhancing the
normalized electron correlation $U/W$, with $W$ being the bandwidth
proportional to the kinetic energy.

Our argument can be checked by fitting band structures of $x=$0, 0.4
and 0.5 to the tight-binding models. We only need to consider the NM
state ( and $U=0$ ) and calculate the sum of the nearest neighbor
hoppings ( absolute values ) among all five d-orbitals around a
specific Fe atom. It serves as an approximate upper limit of the
total kinetic energy $E_K$. Using the
fitted hopping parameters we obtain $E_K=$15.59 and 9.98 eV for
$x=$0 and 0.4, respectively. While for $x=$0.5, we obtain
$E_K=$11.34 or 6.87 eV, corresponding to 3- or 2-coordinated
Fe-sites, respectively. As the ratio of their numbers is 2:1, the
average kinetic energy is 9.85 eV.  For comparison, the DOS at the
NM state when $U=0$ for $x=0, 0.4$, and $0.5$ are plotted in Fig.
\ref{fig_dos_nm}. The Fe-3d bandwidths are roughly 4.8 eV, 3.8 eV,
and 3.5 eV, respectively. Thus, we indeed find a substantial
enhancement of the normalized electron correlation, manifesting the
kinetic energy reduction caused by Fe-vacancies.

\begin{table}[htp]
 \caption{Pressure dependency of crystal structure and
band gap for $x=0.5$ and $U=4.0$ eV.}
 \label{tab_eg_p}
 \begin{tabular}{c|c|c|c|c}
 $P$ (GPa) & 0 & 2 & 4 & 6 \\
 \hline
 a (\AA)     & 5.5507 & 5.4517 & 5.3716 & 5.3079 \\
 b (\AA)     & 11.0342 & 10.8842 & 10.7283 & 10.6177  \\
 c (\AA)     & 13.8027 & 13.5166 & 13.2874 & 13.1102 \\
 $E_{cell}$ (eV) & -187.01 & -186.65 & -185.91 & -158.00 \\
 $E_g$ (meV) & 140 &  80  & 50 & 0 \\
 \end{tabular}
\end{table}

Finally, we remark that our calculations assume homogenous formation
of the Fe-vacancy ordered superstructures. If this is the case in
real materials, the activation gap observed in the $x=0.5$ sample
should be not due to the Anderson localization caused by vacancy
disorder effect. As the Mott (de-)localization is sensitive to
physical pressure which can lead to monotonous bandwidth expansion
by lattice contraction, we expect that the gap dependence on
pressure can be used to test this scenario. The numerical results
for the pressure dependent band gap $E_g$ with fixed $U=4.0$ eV are
listed in TAB. \ref{tab_eg_p} ( $x=0.5$ ). From these results we
expect that the activation gap, which is about 60 meV or well below
150 meV \cite {mhfang_1012} under ambient pressure, can be
completely suppressed by applying pressure up to $\sim $6 GPa and
then the SC may emerge.

\begin{table}[htp]
 \caption{Tight-binding Hamiltonian for TlFe$_{2}$Se$_2$.
Only the nearest neighbor hoppings and on-site energies ( diagonal
line in the brackets ) are shown here. All numbers are in eV.}
 \label{tab_ham}
 \begin{tabular}{c||c|c|c|c|c}
     &  d$_{z^2}$ &  d$_{zx}$ &  d$_{zy}$ &  d$_{x^2-y^2}$ & d$_{xy}$ \\
 \hline\hline
 d$_{z^2}$    &  -0.019 & 0 & 0.156 &  0.303 &  0 \\
     & (4.765) & & & & \\
 \hline
 d$_{zx}$  &   0  &  -0.028 &  0 & 0  & 0.286\\
     &  & (5.084) & & & \\
 \hline
 d$_{zy}$  &  0.156  & 0 & -0.314 & -0.344  & 0 \\
     &  &  & (5.084) & &  \\
 \hline
 d$_{x^2-y^2}$ & -0.303 & 0  & -0.344   & -0.378   & 0.0\\
     &  &  &  & (4.600) & \\
 \hline
 d$_{xy}$  & 0 & 0.286 & 0 & 0 & -0.047 \\
     &  &  &  &  & (5.056) \\
 \end{tabular}
\end{table}

To summarize, we find that for TlFe$_{1.5}$Se$_2$, the electronic
band structure shows a sizable band gap if the short-ranged Coulomb
interaction $U$ beyond LDA/GGA is considered. From the
experimentally observed activation gap, $U$ should be at least 2eV.
The corresponding ground state is then a stripe-like
anti-ferromagnetic Mott insulator. The superstructure of various
Fe-vacancy ordering patterns is important for the stability of the
Mott insulating phase and kinetic energy reduction caused by the
vacancies plays a crucial role for the Mott localization. The
pressure dependence of the gap behavior could be used to test this
possibility.

We would like to express special thanks to M. Fang and Q. Si for
helpful discussions. We are also grateful for the discussions with
G. Cao, H. Wang, Z. Xu, H. Yuan. This work was supported by the
NSFC, the 973 Project of the MOST and the Fundamental Research Funds
for the Central Universities of China (No. 2010QNA3026). All the
calculations were performed at High Performance Computing Center of
Hangzhou Normal University.

Note added: After completing this work, we became aware of two
recent papers by X.W. Yan et al.\cite{zylu_new}, where independent
first-principles studies on the cases of $x=0$ and $x=0.5$ were
reported.

\bibliographystyle{apsrev}
\bibliography{tl122}
\end{document}